\documentclass[preprintnumbers,amsmath,amssymb,floatfix,11pt,prd,twocolumn,superscriptaddress,nofootinbib]{revtex4}
\usepackage[colorlinks=true, pdfstartview=FitV, linkcolor=magenta, citecolor=red, urlcolor=blue]{hyperref}
\usepackage{epsfig,color}
\usepackage{xcolor}
\usepackage{epstopdf}
\usepackage{amssymb}
\usepackage{verbatim}
\usepackage{multirow}
\usepackage{latexsym}
\usepackage{epsfig}
\usepackage{epstopdf}
\usepackage{graphicx}
\usepackage{amssymb}
\usepackage{amsmath}
\usepackage{dcolumn}
\usepackage{bm}
\usepackage{color}
\usepackage{comment}
\usepackage{mathrsfs}
\usepackage{calrsfs}
\DeclareMathAlphabet{\pazocal}{OMS}{zplm}{m}{n}

\begin{document}

\title{Quantum-Corrected $\phi^{4}$ Inflation in Light of ACT Observations}

\author{Jureeporn Yuennan}
\email{jureeporn\_yue@nstru.ac.th}
\affiliation{Faculty of Science and Technology, Nakhon Si Thammarat Rajabhat University, Nakhon Si Thammarat, 80280, Thailand}

\author{Peeravit Koad}
\email{harrykoad@gmail.com}
\affiliation{Informatics Innovation Center of Excellence (IICE) \& School of Informatics, Walailak University, Nakhon Si Thammarat, 80160, Thailand}

\author{Farruh Atamurotov}
\email{atamurotov@yahoo.com}
\affiliation{Urgench State University, Kh. Alimdjan str. 14, Urgench 220100, Uzbekistan}

\author{Phongpichit Channuie}
\email{phongpichit.ch@mail.wu.ac.th}
\affiliation{School of Science \& College of Graduate Studies, Walailak University, Nakhon Si Thammarat, 80160, Thailand}

\date{\today}

\begin{abstract}
Recent measurements from the Atacama Cosmology Telescope (ACT), combined with Planck and DESI data, suggest a scalar spectral index $n_s$ higher than the Planck 2018 baseline, thereby placing conventional attractor-type inflationary models such as Starobinsky $R^2$ and Higgs inflation under increasing tension at the $\gtrsim 2\sigma$ level. In this work, we examine quantum-corrected $\phi^4$ inflation with a non-minimal coupling to gravity. Introducing an anomalous scaling parameter $\gamma$ to capture quantum corrections to the effective potential, we derive analytic expressions for the inflationary observables $n_s$ and $r$. Confronting these predictions with ACT, Planck, and BAO+lensing constraints, we demonstrate that modest values of $\gamma$ can raise $n_s$ into the ACT-preferred range while maintaining a strongly suppressed tensor-to-scalar ratio. For instance, with $N=60$ and $\gamma\simeq 0.006$, the model predicts $n_s\simeq 0.974$ and $r\simeq 0.007$, in excellent agreement with current bounds. We further investigate preheating dynamics, focusing on particle production via parametric resonance in quantum-corrected $\phi^4$ inflation with a non-minimal coupling to gravity. In this scenario, the inflaton $\phi$ couples to an additional scalar $\chi$ through an interaction $g^{2}\phi^{2}\chi^{2}$. In Minkowski spacetime, the resonance dynamics reduce to the Mathieu equation, and we find that broad resonance can be readily achieved, leading to efficient particle production. 
\end{abstract}


\maketitle

\section{Introduction}
Inflation has emerged as a fundamental element of contemporary cosmology, furnishing a persuasive explanation for the flatness, horizon, and monopole puzzles that challenge the standard Big Bang framework. At the same time, it naturally accounts for the generation of primordial fluctuations, which seeded the growth of cosmic structures and manifested as anisotropies in the cosmic microwave background (CMB) \cite{Starobinsky:1980te,Sato:1981qmu,Guth:1980zm,Linde:1981mu,Albrecht:1982wi}. These early fluctuations are typically described through two principal parameters: the scalar spectral index $n_s$, indicating the scale dependence of scalar modes, and the tensor-to-scalar ratio $r$, which gauges the amplitude of primordial gravitational waves relative to scalar perturbations. For a chosen inflationary potential, both observables are usually written as functions of the number of $e$-foldings $N$ between the time modes cross the horizon and the end of inflation, enabling precise theoretical expectations to be matched against observational data, in which the spectral index takes the universal form $n_{s}=1-\tfrac{2}{N}$. This expression is realized in several scenarios, including the $\alpha$-attractor models \cite{Kallosh:2013tua,Kallosh:2013hoa,Kallosh:2013maa,Kallosh:2013yoa,Kallosh:2014rga,Kallosh:2015lwa,Roest:2015qya,Linde:2016uec,Terada:2016nqg,Ueno:2016dim,Odintsov:2016vzz,Akrami:2017cir,Dimopoulos:2017zvq}, the $R^2$ inflationary framework \cite{Starobinsky:1980te}, and Higgs inflation with a strong nonminimal coupling to gravity \cite{Kaiser:1994vs,Bezrukov:2007ep,Bezrukov:2008ej}. Moreover, the models of inflation driven by the composite inflaton fields \cite{Channuie:2011rq}, see reviews \cite{Channuie:2014ysa,Samart:2022pza}, also predict the same universal form. For the benchmark value $N=60$, the prediction $n_s \approx 0.9667$ shows excellent consistency with the {\it Planck} 2018 result, $n_s = 0.9649 \pm 0.0042$ \cite{Planck:2018jri}.
Nevertheless, recent measurements from the Atacama Cosmology Telescope (ACT) \cite{ACT:2025fju, ACT:2025tim}, when combined with other probes, suggest a noticeably larger value of the scalar spectral index than that inferred by {\it Planck} alone. A combined analysis of ACT and {\it Planck} (P-ACT) reports $n_s = 0.9709 \pm 0.0038$, and incorporating CMB lensing along with baryon acoustic oscillation data from DESI \cite{DESI:2024uvr, DESI:2024mwx} (P-ACT-LB) further raises the estimate to $n_s = 0.9743 \pm 0.0034$ \cite{ACT:2025fju, ACT:2025tim}. 

These updated results place universal attractors under increasing pressure, excluding them at roughly the $2\sigma$ level and thereby creating a serious challenge for many inflationary scenarios that predict this behavior. In Ref.\cite{ACT:2025fju}, the authors specifically point out that the P-ACT-LB bounds on $n_s$ place the Starobinsky model under tension at the level of $\gtrsim 2\sigma$. Such a conclusion is both striking and somewhat surprising, standing in clear contrast to earlier findings. A variety of strategies have been put forward to alleviate the discrepancy between recent observational results and predictions of standard inflationary scenarios. Among them are relaxing the strong-coupling assumption in nonminimally coupled models of the form $\xi f(\phi)R$ \cite{Kallosh:2025rni,Gao:2025onc}, incorporating the effects of reheating dynamics \cite{Haque:2025uis,Zharov:2025evb,Haque:2025uri,Drees:2025ngb,Ballardini:2024ado,Ye:2025idn,SidikRisdianto:2025qvk,Chakraborty:2025oyj}, and exploring alternative classes of inflationary constructions \cite{Wolf:2025ecy,He:2025bli,Gialamas:2025kef,Frob:2025sfq,Brahma:2025dio,Berera:2025vsu,Aoki:2025wld,Dioguardi:2025mpp,Salvio:2025izr,Gialamas:2025ofz,Gao:2025viy,Peng:2025bws,Yi:2025dms,Pallis:2025nrv,Katsoulas:2025mcu,Byrnes:2025kit,Maity:2025czp,Addazi:2025qra,Mondal:2025kur,Odintsov:2025wai,Yogesh:2025wak,Zahoor:2025nuq}. A comprehensive summary of these developments can be found in Ref.~\cite{Kallosh:2025ijd}.

In this work, we study whether the quantum corrections on the naive $\phi^4$-inflation, can provide a better fit to the latest observational data. We consider the scenario where the inflaton field non-minimally couples to gravity \cite{Joergensen:2014rya}. The structure in the present work is the following: Sec.~\ref{sec1}, we begin by reviewing the quantum corrections $\phi^4$-inflation (QC). After moving from the Jordan to the Einstein frame,
as described in Sec.~\ref{sec1}, the resulting action contains a non-canonical kinetic term for the Higgs field. Along with the standard fashion, the kinetic term
for the inflaton can be normalized by the field
redefinition. Having canonically normalized both the gravity and Higgs kinetic terms, we come up with the effective potential in the Einstein frame. Consequently, the expressions for the scalar spectral index and tensor-to-scalar ratio can be straightforwardly obtained. In Sec.\ref{sec2}, we confront the predictions with recent observational data. In Sec.(\ref{pre}), we study parametric resonances of model when the inflaton field $\phi$ coupled to another scalar field $\chi$ with the interaction term $g^{2}\phi^{2}\chi^{2}$.  Finally, in Sec.~\ref{con}, we summarize our results.

\section{Quantum Corrected Inflation}\label{sec1}
The origin of inflation remains a fundamental challenge in cosmology. Scalar fields are often introduced to model inflationary dynamics, with a variety of proposals in the literature. In general, quantum corrections modify the classical scalar potential \cite{Weinberg:1973am,Gildener:1976ih}. To capture these corrections in a simple framework, we parametrize the inflaton potential as \cite{Joergensen:2014rya}
\begin{equation}
V_{\text{eff}} = \lambda \phi^{4} \left( \frac{\phi}{\Lambda} \right)^{4\gamma}, 
\end{equation}
where $\Lambda$ is an energy scale and $\gamma$ encodes anomalous scaling effects. In the non-minimally coupled case, BICEP2 results on primordial tensor modes constrain $\gamma$ to the range $0.08 \leq \gamma \leq 0.12$ at two-sigma confidence level \cite{Joergensen:2014rya}. Regardless of the fate of these data, it is crucial to investigate whether quantum-corrected potentials generate observable tensor modes. Interestingly, for large tensor signals, predictions are largely insensitive to the total number of e-folds. Examples include Higgs inflation and composite inflaton models. In our analysts below, we follow Ref.\cite{Joergensen:2014rya} and consider the Jordan-frame action
\begin{eqnarray}
&&S_J = \int d^4x\sqrt{-g}\Big[-\frac{1}{2}(M^2+\xi \phi^2)R \nonumber\\&& \quad\quad\quad\quad\quad\quad\quad +g^{\mu\nu}\partial_\mu\phi\partial_\nu\phi - V_{\text{eff}}(\phi) \Big].
\end{eqnarray}
In the Jordan frame, the scalar background typically contributes to the effective Planck mass as $M^{2}_{p}=\left\langle M^{2}+\xi\phi^2 \right\rangle$. In the present case, however, since $\left\langle \phi^2 \right\rangle
=0$, we can safely identify $M$ with $M_{p}$. A conformal transformation
\begin{equation}
g_{\mu\nu} \to \tilde{g}_{\mu\nu} = \Omega(\phi)^2 g_{\mu\nu}, 
\quad \Omega(\phi)^2 = 1 + \frac{\xi \phi^2}{M_p^2},
\end{equation}
yields the Einstein-frame action
\begin{widetext}
\begin{equation}
S_E = \int d^4x \, \sqrt{-g} \left[ -\frac{1}{2}M_p^2 R + \Omega^{-2}(1+3M_p^2\Omega'^2) g^{\mu\nu}\partial_\mu\phi\partial_\nu\phi - \Omega^{-4}V(\phi)\right].
\end{equation}
\end{widetext}
Defining a canonical field $\chi$ via
\begin{equation}
\frac{1}{2}\left(\frac{d\chi}{d\phi}\right)^2 = \frac{M_p^2\left(M_p^2+(1+3\xi)\xi \phi^2\right)}{(M_p^2+\xi\phi^2)^2},
\end{equation}
the action becomes
\begin{eqnarray}
&&S_E = \int d^4x \, \sqrt{-g}\Big[-\frac{1}{2}M_p^2 R + \frac{1}{2} g^{\mu\nu}\partial_\mu \chi \partial_\nu \chi \nonumber\\&& \quad\quad\quad\quad\quad\quad\quad\quad - U(\chi)\Big],
\end{eqnarray}
with
\begin{equation}
U(\chi) = \Omega^{-4} V_{\text{eff}}(\phi(\chi)).
\end{equation}
In the large-field limit $\phi \gg M_P/\sqrt{\xi}$, one obtains
\begin{equation}
\chi \simeq \kappa M_p \ln\!\left( \frac{\sqrt{\xi}\phi}{M_p} \right), \quad
\kappa \equiv \sqrt{\tfrac{2}{\xi}+6}.
\end{equation}
We consider the regime $\xi \gg 1$, since values of order $\xi \sim 10^4$ are typically required to reproduce the correct amplitude of density perturbations. This behavior is a generic feature of non-minimally coupled single-field inflationary theories \cite{Bezrukov:2007ep,Channuie:2011rq,Channuie:2012bv,Lee:2014spa,Cook:2014dga,Park:2008hz}. Although a smaller $\xi$ can, in principle, be achieved, it would necessitate an exceptionally small value of $\lambda$, as pointed out in \cite{Hamada:2014iga}. The Einstein-frame potential becomes
\begin{eqnarray}
U(\chi) &=& \frac{\lambda M_p^4}{\xi^2} \left( 1 + e^{-\tfrac{2\chi}{\kappa M_p}} \right)^{-2} 
\left(\frac{M_p}{\sqrt{\xi}\Lambda}\right)^{4\gamma} \times\nonumber\\&&\quad\quad\quad\times
\exp\!\left(\frac{4\gamma\chi}{\kappa M_p}\right).
\end{eqnarray}
The slow-roll parameters are
\begin{eqnarray}
\epsilon &=&\frac{M^{2}_{p}}{2}\Big(\frac{dU/d\chi}{U}\Big)^{2}\nonumber\\&=& \frac{8M_p^4}{\kappa^2 \xi^2 \phi^4} + \frac{16M_p^2}{\kappa^2 \xi \phi^2}\gamma + \frac{8}{\kappa^2}\gamma^2+ \mathcal{O}(\gamma^3),\\
\eta &=&\frac{M^{2}_{p}}{2}\Big(\frac{d^{2}U/d\chi^{2}}{U}\Big) \nonumber\\&=& \frac{8}{\kappa^2}\left( -\frac{M_p^2}{\xi \phi^2} + \frac{M_p^4}{\xi^2 \phi^4} + \frac{3M_p^2}{\xi \phi^2}\gamma + 2\gamma^2 \right)\nonumber\\&+& \mathcal{O}(\gamma^3).
\end{eqnarray}
Inflation ends when $\epsilon(\phi_{\text{end}}) = 1$, leading to
\begin{eqnarray}
\phi_{\text{end}} &\simeq& \frac{2M_p}{\sqrt{\xi}}\frac{1}{\sqrt{2\kappa-4\gamma}} 
\nonumber\\&=& \left(1.07+0.62\gamma\right)\frac{M_p}{\sqrt{\xi}} + \mathcal{O}(\gamma^2).
\end{eqnarray}
A universal bound of $\gamma$ has been determined to obtain $\gamma < \tfrac{\sqrt{3}}{2}$ \cite{Joergensen:2014rya}. The number of e-foldings is
\begin{eqnarray}
N &=&\frac{1}{M_p^2}\int^{\chi_{*}}_{\chi_{\rm end}}\frac{U}{dU/d\chi}d\chi\nonumber\\&&=\frac{1}{M_p^2}\int^{\phi_{*}}_{\phi_{\rm end}}\frac{U}{dU/d\phi}\left(\frac{d\chi}{d\phi}\right)^2d\phi\nonumber\\&&= \frac{\kappa^2}{8\gamma} \ln\!\left[ 1 + \gamma\frac{\xi \phi^2}{M_p^2} \right]_{\phi_{\text{end}}}^{\phi_*}.
\end{eqnarray}
The horizon-crossing field value is then
\begin{eqnarray}
\phi_{*} &\simeq& \sqrt{\frac{1}{\gamma}\left(\exp\!\left(\frac{8\gamma N}{\kappa^2}\right)-1\right)} \frac{M_p}{\sqrt{\xi}}\nonumber\\&\simeq &\bigg(1.15 \sqrt{N}+0.38 N^{3/2} \gamma \nonumber\\&&\quad+0.11 N^{5/2} \gamma^2\bigg)\frac{M_{p}}{\sqrt{2}}.
\end{eqnarray}
We notice that when $\gamma\rightarrow 0$ and $N=60$, the value approaches those of $\phi^{4}$-theory. Consequently, the expressions for the scalar spectral index and tensor-to-scalar ratio take the form:
\begin{eqnarray}
n_s &=& 1-6\epsilon+2\eta \simeq 1-\frac{2}{N}+\frac{4}{3}\gamma-\frac{8}{27}N\gamma^2\,, \label{nss}\\
r &=& 16\epsilon \simeq \frac{12}{N^2}+\frac{16}{N}\gamma+\frac{80}{9}\gamma^2\,.\label{rr}
\end{eqnarray}
It is evident that when $\gamma = 0$, the predictions coincide with those of many inflationary models (see, e.g., the Higgs and Higgs-like ~\cite{Kaiser:1994vs,Bezrukov:2007ep} and composite inflaton \cite{Channuie:2011rq}). In the presence of the $\gamma$ parameter, the predictions exhibit a tendency toward larger values for the modest values of $\gamma$.

\section{Confront with Observations}\label{sec2}
In combination with baryon acoustic oscillation (BAO \cite{eBOSS:2020yzd}) and CMB lensing \cite{Planck:2018lbu} data, the authors of Ref.\cite{Tristram:2021tvh} obtained an improved upper limit of $r<0.032$ (95\% C.L.); while allowing for a slight relaxation of the bound $r < 0.038$ (95\% C.L.) reported by P-ACT-LB-BK18 \cite{ACT:2025tim}. This allow us to determine an upper limit of $\gamma$ from Eq.(\ref{rr}) to obtian
\begin{eqnarray}
\gamma < -\frac{0.9}{N}+0.015 \sqrt{\frac{19N^{2}-2400}{N^2}}\,,
\end{eqnarray}
for which $N=60$, we get $\gamma <0.0492$. To produce $n_s=0.9743$, from Eq.(\ref{nss}) we come up with
\begin{eqnarray}
\gamma \rightarrow 0.0059762, \quad \gamma \rightarrow 0.0690238\,,
\end{eqnarray}
where the latter case can be ignored. The inclusion of P-ACT data shows a preference for a slightly higher value of $n_s$, indicated by the green contour. For $\gamma = 0$, the predictions overlap with those of the Starobinsky $R^2$ framework and the Higgs or Higgs-like models. However, in the interval $50 < N < 60$, these scenarios are in tension with the P-ACT-LB measurement of $n_s$ at a confidence level of $\gtrsim 2\sigma$.
\begin{figure}
    \centering
    \includegraphics[width=1\linewidth]{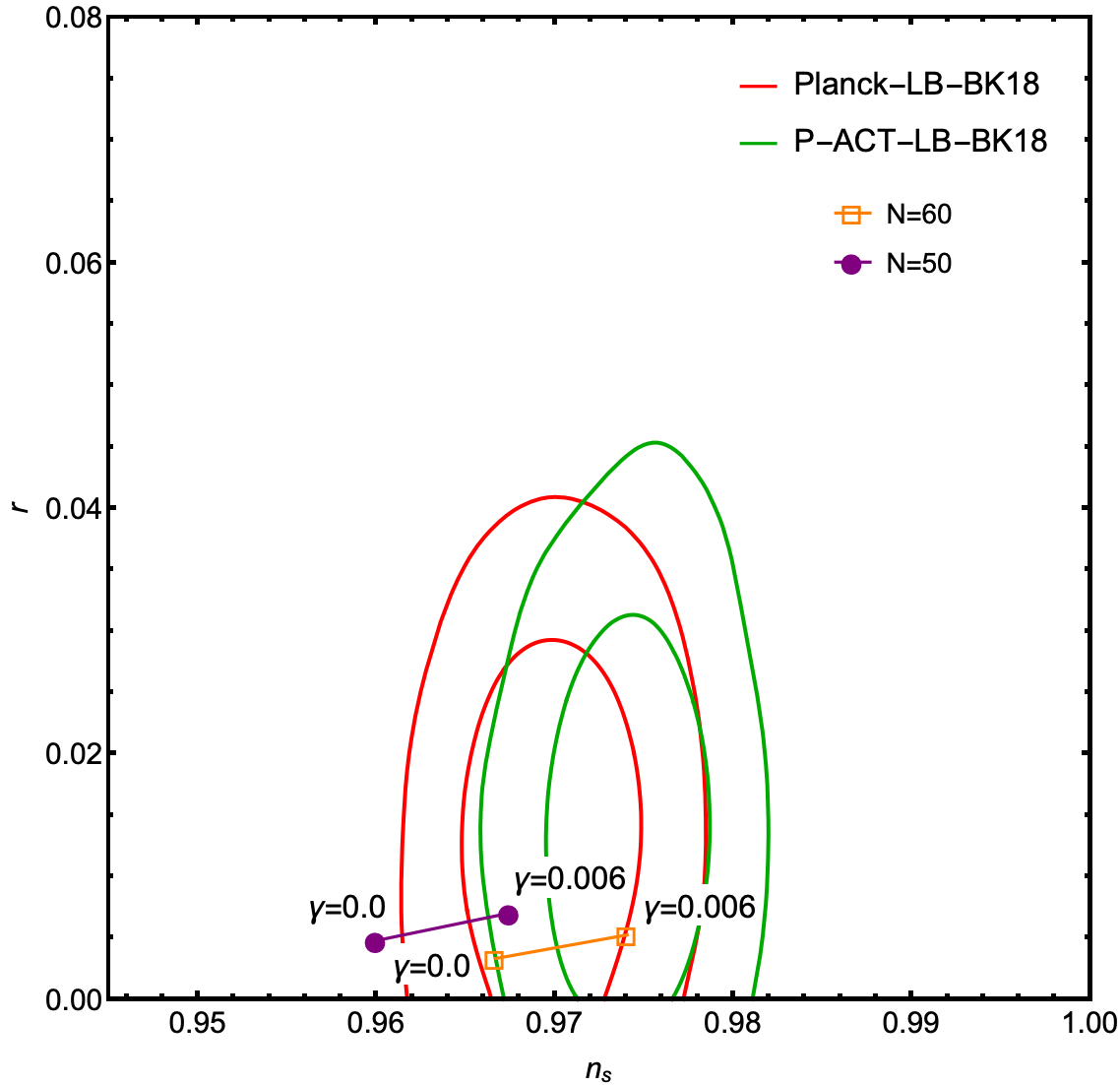}
    \caption{Constraints on the scalar and tensor primordial power spectra at $k_{*}= 0.05$\,Mpc$^{-1}$, shown in the $r-n_{s}$
parameter space. The bounds on $r$ are primarily determined by the BK18 observations, whereas the limits on $n_s$ are set by Planck (red) and P-ACT (green) data. Circles and squares represent predictions from quantum-corrected potentials for $50 < N < 60$ and $0 < \gamma < 0.006$. For $\gamma = 0$, the results coincide with those of the Starobinsky $R^2$ model, as well as Higgs and Higgs-like scenarios. However, within the range $50 < N < 60$, the P-ACT-LB determination of $n_s$ excludes these models at a significance level of at least $2\sigma$.}
    \label{fig1}
\end{figure}

From Fig.~\ref{fig1}, we observe that quantum corrections to the naive $\phi^4$-inflation scenario yield an improved agreement with the latest observational data, particularly when the inflaton field is allowed to couple non-minimally to gravity. In the limit $\gamma=0$, our framework reproduces the predictions of the Starobinsky $R^2$ model, as well as those of Higgs and Higgs-like inflationary scenarios. More specifically, for $\gamma=0.006$ and $N=60$, the model predicts $n_{s}=0.974$ and $r=0.007$, values that are in excellent agreement with the most recent observations.

\begin{figure}
    \centering
    \includegraphics[width=1\linewidth]{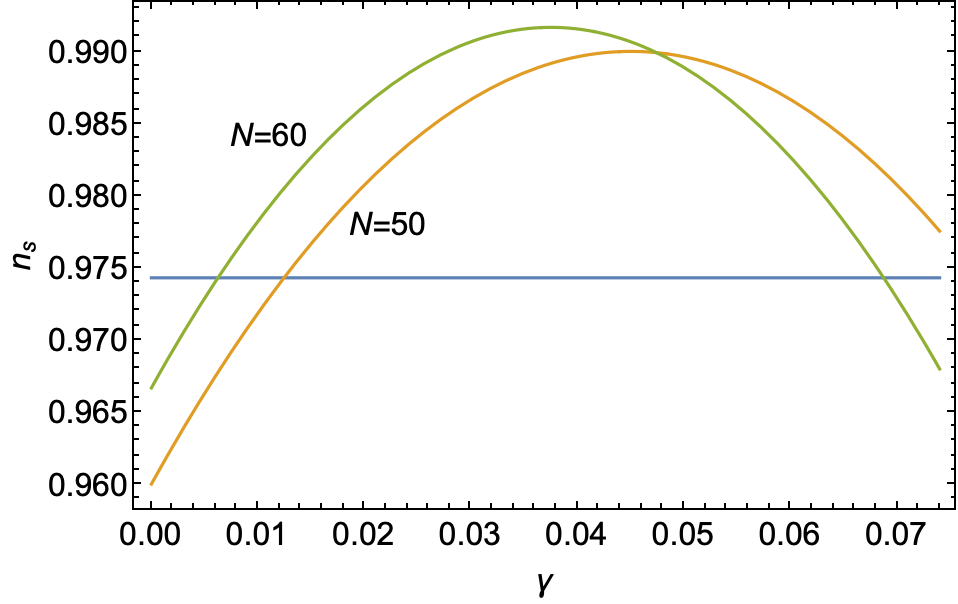}
    \caption{We plot $n_{s}$ vs $\gamma$ for $N=60$ (green) and $N=50$ (orange). The horizontal line denotes $n_s=0.9743$. It is noticed that to obtain the same $n_s$, the model needs $\gamma_{N=50}>\gamma_{N=60}$ for $\gamma<0.047$.}
    \label{fig2}
\end{figure}

The curves Fig.\ref{fig2} indicate that $n_s$ grows with $\gamma$ at small values, reaches a maximum around $\gamma \sim 0.04$-$0.05$, and then decreases, resulting in the concave--down behavior for both $N=60$ (green) and $N=50$ (orange). At $\gamma=0$, the predictions are $n_s \simeq 0.9667$ for $N=60$ and $n_s \simeq 0.9600$ for $N=50$, implying that a positive $\gamma$ is required to reproduce the reference value $n_s=0.9743$.

\section{Parametric resonances in QC inflation}\label{pre}
In this study, we investigate the preheating phase within a composite inflationary framework that incorporates a non-minimal interaction between matter and gravity. Models of this type have been widely explored in the literature, including Higgs inflation~\cite{Bezrukov:2007ep} and the pure $\lambda\phi^{4}$ theory~\cite{Futamase:1987ua,Mahajan:2013kra,Steinwachs:2013tr} including a composite scenario \cite{Channuie:2016xmq}. Following the formalism outlined in~\cite{Kaiser:2010ps}, we consider the two-field setup, which itself has been frequently analyzed in earlier works~\cite{Bassett:1997az,Tsujikawa:2002nf,Amin:2014eta}. The impact of preheating in the more general multifield case has also been examined, see e.g.~\cite{DeCross:2015uza}.  

For our present purpose, we adopt the following four-dimensional Jordan (J) frame action:
\begin{eqnarray}
\mathcal{S}_{\rm J}&&=\int d^{4}x \sqrt{-g}\Big[- f(\Phi^{i})R \nonumber\\
&&\,\quad\quad\quad\quad+\frac{1}{2}\delta_{ij}g^{\mu\nu}\nabla_{\mu}\Phi^{i}\nabla_{\nu}\Phi^{j} -V(\Phi^{i}) \Big]\,, \label{actionset}
\end{eqnarray}
where $i=1,2$ and $\Phi^{i}=(\phi,\chi)$, with $\chi$ denoting an additional scalar degree of freedom. We take the non-minimal coupling function in the form
\begin{eqnarray}
f(\Phi^{i}) = \frac{1}{2}\left(M^{2}_{0} + \xi_{\Phi^{i}}(\Phi^{i})^{2}\right)\,, \label{fP}
\end{eqnarray}
where $M_{0}$ is identified with the reduced Planck mass $M_{\rm p}$, while $\xi_{\Phi^{i}}$ represents the non-minimal couplings of the fields to the curvature. To bring the gravitational sector into canonical Einstein–Hilbert form, we perform a conformal transformation $\tilde{g}_{\mu\nu} = \Omega^{2}(x)g_{\mu\nu}$, with the conformal factor defined as
\begin{eqnarray}
\Omega^{2}(x) = \frac{2}{M^{2}_{p}}f(\Phi^{i}(x))\,. \label{con}
\end{eqnarray}
This transformation removes the explicit non-minimal terms, leading to the action in the Einstein frame~\cite{Kaiser:2010ps}:
\begin{eqnarray}
\mathcal{S}_{\rm E}&=&\int d^{4}x \sqrt{-g}\Big[- \frac{M^{2}_{p}}{2}R \nonumber\\
&&\,+ \frac{1}{2}{\cal G}_{ij}g^{\mu\nu}\nabla_{\mu}\Phi^{i}\nabla_{\nu}\Phi^{j} -{\cal U}(\Phi^{i})\Big]\,, \label{ef}
\end{eqnarray}
with ${\cal U}(\Phi^{i})\equiv V(\Phi^{i})/\Omega^{4}$ and the field-space metric
\begin{eqnarray}
{\cal G}_{ij} = \frac{M^{2}_{p}}{2f}\delta_{ij} + \frac{3M^{2}_{p}}{2f^{2}}f_{,i}f_{,j}\,, \label{Gij}
\end{eqnarray}
where $f_{,i}=\partial f/\partial\Phi^{i}$. Explicitly, for the fields $(\phi,\chi)$ one obtains
\begin{eqnarray}
{\cal G}_{\phi\phi} &=& \frac{M^{2}_{p}}{2f}\left(1+\frac{3\xi_{\phi}\phi^{2}}{f}\right)\,, \label{pp}\\
{\cal G}_{\phi\chi} &=& {\cal G}_{\chi\phi} = \frac{M^{2}_{p}}{2f}\left(\frac{3\xi_{\phi}\xi_{\chi}\phi\chi}{f}\right)\,, \label{pc}\\
{\cal G}_{\chi\chi} &=& \frac{M^{2}_{p}}{2f}\left(1+\frac{3\xi_{\chi}\chi^{2}}{f}\right)\,.\label{cc}
\end{eqnarray}
The full action then takes the form
\begin{eqnarray}
\mathcal{S}_{\rm E}&=&\int d^{4}x \sqrt{-g}\Big[- \frac{M^{2}_{\rm p}}{2}R \nonumber\\
&&\,+ \frac{1}{2}\frac{M^{2}_{p}}{2f}\left(1+\frac{3\xi_{\phi}\phi^{2}}{f}\right)g^{\mu\nu}\nabla_{\mu}\phi\nabla_{\nu}\phi \nonumber\\
&&\,+ \frac{1}{2}\frac{M^{2}_{p}}{2f}\left(\frac{3\xi_{\phi}\xi_{\chi}\phi\chi}{f}\right)g^{\mu\nu}\nabla_{\mu}\phi\nabla_{\nu}\chi \nonumber\\
&&\,+ \frac{1}{2}\frac{M^{2}_{p}}{2f}\left(1+\frac{3\xi_{\chi}\chi^{2}}{f}\right)g^{\mu\nu}\nabla_{\mu}\chi\nabla_{\nu}\chi \nonumber\\
&&\,-{\cal U}(\phi,\chi)\Big]\,. \label{actionE2}
\end{eqnarray}

We emphasize that in this two-field construction, each field couples differently to curvature. Consequently, no conformal redefinition exists that simultaneously casts both the gravitational and the scalar kinetic terms into fully canonical form. To proceed, we assume that only $\phi$ is non-minimally coupled, setting $\xi_{\phi}=\xi$, while $\chi$ remains minimally coupled ($\xi_{\chi}=0$). The resulting action then reads
\begin{eqnarray}
\mathcal{S}_{\rm E}&=&\int d^{4}x \sqrt{-g}\Big[- \frac{M^{2}_{p}}{2}R \nonumber\\
&&\,+ \frac{1}{2}\frac{M^{2}_{p}}{2F}\left(1+\frac{3\xi\phi^{2}}{F}\right)g^{\mu\nu}\nabla_{\mu}\phi\nabla_{\nu}\phi \nonumber\\
&&\,+ \frac{1}{2}\frac{M^{2}_{p}}{2F}g^{\mu\nu}\nabla_{\mu}\chi\nabla_{\nu}\chi \nonumber\\
&&\,-{\cal U}(\phi,\chi)\Big]\,, \label{actionE3}
\end{eqnarray}
where $F \equiv f|_{\xi_{\chi}=0}$. Canonical normalization can be restored through rescaled fields $\hat{\phi}(\phi,\chi)$ and $\hat{\chi}(\phi,\chi)$ defined via
\begin{eqnarray}
\frac{\partial\hat{\phi}}{\partial\phi} &=& \sqrt{\frac{M^{2}_{p}}{2F}\left(1+\frac{3\xi\phi^{2}}{F}\right)}\,,\nonumber\\
\frac{\partial\hat{\chi}}{\partial\chi} &=& \sqrt{\frac{M^{2}_{p}}{2F}}\,. \label{resca12}
\end{eqnarray}
In terms of the rescaled fields, the action becomes
\begin{eqnarray}
\mathcal{S}_{\rm E}&=&\int d^{4}x \sqrt{-g}\Big[ - \frac{M^{2}_{p}}{2}R +\frac{1}{2}(\nabla\hat{\phi})^{2} + \frac{1}{2}(\nabla\hat{\chi})^{2} \nonumber\\
&&\quad\quad\quad\quad\quad -{\cal U}\!\left(\hat{\phi}(\phi,\chi),\hat{\chi}(\phi,\chi)\right)\Big]\,. \label{actionEi}
\end{eqnarray}
After the conformal (Weyl) transformation, the potential in terms of the rescaled fields $U(\hat{\phi},\hat{\chi})$ takes the form
\begin{eqnarray}
U\!\left(\hat{\phi},\hat{\chi}\right) &=& \exp\!\left(-2\sqrt{\tfrac{2}{3}}\tfrac{\hat{\phi}}{M_{p}}\right)\Big[\lambda \phi^{4} \Big( \frac{\phi}{\Lambda} \Big)^{4\gamma} \nonumber\\
&&\quad+\frac{1}{2}g^{2}\phi^{2}\chi^{2} - \frac{1}{2}m^{2}_{\chi}\chi^{2} \Big]\,. \label{actionEPott}
\end{eqnarray}
Relabelling the canonically normalized fields ($\hat{\phi}\to\phi$, $\hat{\chi}\to\chi$), the potential can be simply expanded up to the first-order of a parameter $\gamma$ to obtain
\begin{widetext}
\begin{eqnarray}
U(\phi,\chi) &\simeq& \frac{\lambda  M_{p}^4}{\xi^2}\left(1-e^{-\frac{\sqrt{\frac{2}{3}} \phi }{M_{p}}}\right)^2+\frac{g^2 M_{p}^2}{2 \xi } e^{-\frac{2 \sqrt{\frac{2}{3}} \phi }{M_{p}}} \left(1-e^{\frac{\sqrt{\frac{2}{3}} \phi }{M_{p}}}\right)\chi^2\nonumber\\&&- \frac{1}{2}m^{2}_{\chi}e^{-2\sqrt{2/3}\phi/M_{p}}\chi^{2}+\frac{4 \lambda M_{p}^4}{\xi ^2}\left(1-e^{-\frac{ \sqrt{\frac{2}{3}} \phi }{M_{p}}}\right)^2 \log \left(\sqrt{1-e^{\frac{\sqrt{\frac{2}{3}} \phi }{M_{p}}}}\frac{M_{p}}{\sqrt{\xi }}\right)\gamma\,, \label{actionEPot}
\end{eqnarray}
\end{widetext}
where we have ignored the inflaton mass contribution. The Klein–Gordon equation for the inflaton is then
\begin{eqnarray}
\ddot{\phi}+3H\dot{\phi}+ {\cal M}^{2}\phi = 0, 
\,\, {\cal M}^{2}\equiv \frac{4 \lambda  M_{p}^2}{3 \xi^2}(1+\gamma)\,. \label{ST_2.1}
\end{eqnarray}

As argued in~\cite{GarciaBellido:2008ab}, the backreaction of $\chi$ quanta on the inflaton background becomes important only once the occupation numbers grow significantly; in the early stage of resonance we neglect this effect. For later analysis we shall compare with~\cite{Kofman:1997yn}. To solve Eq.~(\ref{ST_2.1}), we assume a power-law expansion $a\propto t^{p}$, giving
\begin{eqnarray}
t^{2}\ddot{\phi}+3pt\dot{\phi} + t^{2}{\cal M}^{2}\phi = 0\,. \label{ST_2.11}
\end{eqnarray}
The general solution can be expressed through Bessel functions:
\begin{eqnarray}
\phi(t) = \frac{1}{({\cal M}t)^{\nu}}\Big[A J_{+\nu}({\cal M}t) + B J_{-\nu}({\cal M}t)\Big]\,, \label{ST_2.111}
\end{eqnarray}
with $\nu=(3p-1)/2$. Physical considerations require discarding the divergent branch, leaving
\begin{eqnarray}
\phi(t) = A\,({\cal M}t)^{-\tfrac{3p-1}{2}}J_{\tfrac{3p-1}{2}}({\cal M}t)\,. \label{ST_2.1111}
\end{eqnarray}
For ${\cal M}t\gg 1$, this admits an approximate oscillatory form
\begin{eqnarray}
\phi(t) \simeq A({\cal M}t)^{-\tfrac{3p}{2}}
\cos\!\left({\cal M}(t-t_{\rm os}) - \tfrac{3p\pi}{4}\right)\,, \label{ST_2.12}
\end{eqnarray}
\begin{figure}[t]
\begin{center}		
\includegraphics[width=1\linewidth]{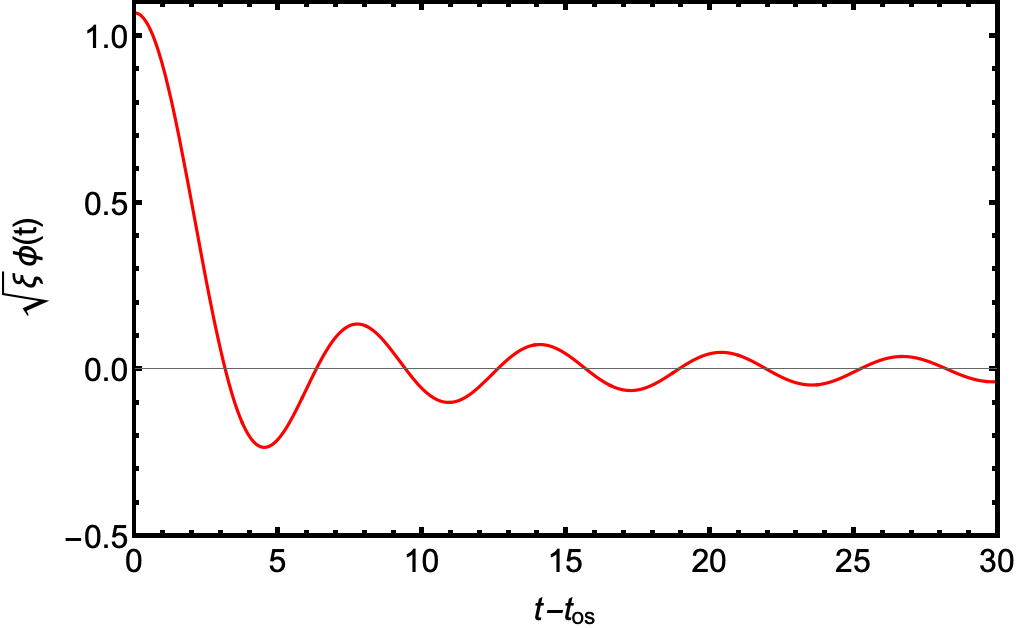}
\caption{We plot the approximate solution of the field $\sqrt{\xi}\phi(t)$ as given in Eq.(\ref{ST_2.21}). The value of the scalar field here is measured in units of $M_{p}$ and time is measured in units of ${\cal M}^{-1}$. \label{phios}}
\end{center}
\end{figure}
where $t_{\rm os}$ marks the onset of oscillations. Choosing initial amplitude $\phi_{\rm end}=\left(1.07+0.62\gamma\right)\frac{M_p}{\sqrt{\xi}} + \mathcal{O}(\gamma^2)$ at the end of inflation, one obtains the early-time solution for $p=2/3$ (matter-like behavior):
\begin{eqnarray}
\phi(t) = \frac{\phi_{\rm end}}{{\cal M}(t-t_{\rm os})}\sin\!\left({\cal M}(t-t_{\rm os})\right)\,. \label{ST_2.21}
\end{eqnarray}
The field $\chi$ obeys
\begin{eqnarray}
\ddot{\chi}+3H\dot{\chi} - \frac{1}{a^{2}}\nabla^{2}\chi +\Big[2m^{2}_{\chi} + {\cal A}^{2}\phi\Big]\chi = 0\,.  \label{ST_2.2}
\end{eqnarray}
where ${\cal A}^{2}\equiv \frac{\sqrt{\frac{2}{3}}}{M_{p} \xi }\left(g^2 M_{p}^2+2 m_{\chi}^2 \xi \right)$. Expanding $\chi$ into Fourier modes,
\begin{eqnarray}
\chi(t,{\bf x})=\!\int\!\!\left(a_{k}\chi_{k}(t)e^{-i{\bf k}\cdot{\bf x}}+a^{\dagger}_{k}\chi^{*}_{k}(t)e^{i{\bf k}\cdot{\bf x}}\right)\!d^{3}{\bf k}, \label{mospace}
\end{eqnarray}
leads to
\begin{eqnarray}
\ddot{\chi}_{k}+3H\dot{\chi}_{k} +\Big[\frac{k^{2}}{a^{2}}+2m^{2}_{\chi}+ {\cal A}^{2}\phi\Big]\chi_{k}=0\,. \label{Pot211}
\end{eqnarray}
Defining $Y_{k}=a^{3/2}\chi_{k}$ gives
\begin{eqnarray}
\ddot{Y}_{k}+\omega_{k}^{2}Y_{k}=0\,, \label{Pot2111}
\end{eqnarray}
with a time-dependent frequency
\begin{eqnarray}
\omega^{2}_{k}=\frac{k^{2}}{a^{2}}+2m^{2}_{\chi}+\Big[{\cal A}^{2}\tfrac{\phi_{\rm end}}{T(t)}\sin(T(t))\Big]\!, \label{Pot20}
\end{eqnarray}
where $T(t)={\cal M}(t-t_{\rm os})$. In Minkowski space ($a=1$), Eq.~(\ref{Pot2111}) reduces to the Mathieu equation:
\begin{eqnarray}
\frac{d^{2}Y_{k}}{dz^{2}}+\left(A_{k}-2q\cos(2z)\right)Y_{k}=0\,, \label{Mathieu}
\end{eqnarray}
with
\begin{eqnarray}
A_{k}&=&\frac{4}{{\cal M}^{2}}\left(k^{2}+2m^{2}_{\chi}\right),\\ q&=&\sqrt{\frac{3}{2}}\frac{1}{(1+\gamma)\beta}\bigg(g^{2}+2\xi \tfrac{m^{2}_{\chi}}{M^{2}_{p}}\bigg)\frac{\phi_{\rm end}}{T(t)}\,, \label{Matheiu2}
\end{eqnarray}
and $\beta=\lambda M_{\rm P}/\xi$. The Mathieu system exhibits instability bands with Floquet exponent $\mu_{k}$, leading to exponential growth $Y_{k}\propto \exp(\mu_{k}z)$ and thus particle production with $n_{k}\propto \exp(2\mu_{k}z)$. Efficient production requires broad resonance ($q\gg 1$), which in our setup demands
\begin{eqnarray}
q\simeq \frac{\sqrt{\frac{3}{2}} \sqrt{\xi }}{(\gamma +1) \lambda }\Big(g^2+\frac{2 m^2_{\chi} \xi }{M_{p}^2}\Big)\gg 1\,, 
\end{eqnarray}
implying that
\begin{eqnarray}
\Bigg(\frac{m_{\chi}}{M_{p}}\Bigg)^{2}\gg (1+\gamma) \sqrt{\frac{\lambda ^2}{6 \xi ^3}}-\frac{g^2}{2 \xi }\,,
\end{eqnarray}
with $(\gamma +1) \sqrt{\tfrac{\lambda ^2}{6 \xi ^3}}>\tfrac{g^2}{2 \xi }$. The number density of produced particles is computed as~\cite{Kofman:1997yn}
\begin{eqnarray}
n_{k}(z)=\frac{\omega_{k}}{2}\left(\frac{|\dot{Y}_{k}|^{2}}{\omega_{k}^{2}}+|Y_{k}|^{2}\right)-\frac{1}{2}\,. \label{growth}
\end{eqnarray}
For example, taking $m \to 10^{-3}M_{p}\,, \xi \to 10^4\,, g\to 10^{-2}\,, \lambda \to 2\times 10^{-2}$, we obtain $q\sim 100$. For representative parameters ($k=5{\cal M}=m_{\chi}$, $q\simeq 10^{2}$), the mode functions exhibit exponential amplification, and the comoving particle number density grows as $\ln n_{k}\simeq 2\mu_{k}z$. Resonance occurs when the inflaton oscillations drive $\phi(t)$ across zero, triggering bursts of particle production at each crossing.  

In the simplest inflationary frameworks---including the one under consideration---the Hubble parameter at the end of inflation is generally of the same order as the effective inflaton mass, ${\cal M}$, though slightly smaller. To make this concrete, we can evaluate the Hubble scale during the first oscillation. As illustrated in Fig.(\ref{phios}), the field amplitude $\phi(t)$ decreases to roughly one-fourth of the reduced Planck mass, $M_{p}$, within the first oscillation. During this early stage, the kinetic and potential energies of the field are expected to be approximately comparable, giving an energy density of 
\begin{equation}
\rho \sim {\cal M}^{2}\phi^{2} \sim \tfrac{1}{16}{\cal M}^{2}M_{p}^{2}.
\end{equation}
Substituting this into the Friedmann relation yields an estimate for the Hubble rate,
\begin{equation}
H = \sqrt{\frac{\rho}{3M_{p}^{2}}} \sim \frac{{\cal M}}{\sqrt{48}} \simeq 0.14\,{\cal M}.
\end{equation}
This ratio, $H/{\cal M} \sim 0.14$, agrees with the results obtained through a more rigorous analysis presented in Ref.~\cite{DeCross:2015uza,Channuie:2016xmq}, Fig.~(8), in the limit $\xi_{\phi} \gg 1$.

\section{Conclusion}\label{con}
In summary, our analysis demonstrates that quantum corrections to the inflaton potential can significantly improve the agreement between theoretical predictions and the most recent ACT and Planck observations. We showed that the inclusion of the anomalous scaling parameter $\gamma$ allows the model to shift $n_s$ toward the higher values favored by P-ACT-LB, while maintaining a suppressed tensor-to-scalar ratio $r$ well below current observational limits. This places the quantum-corrected scenario in better alignment with the data compared to the minimal Starobinsky $R^2$ and Higgs-like models, which are increasingly disfavored at the $\gtrsim 2\sigma$ level. Moreover, the degeneracy between $(N,\gamma)$ for a fixed $n_s$ highlights the need for complementary probes, such as measurements of $r$, to fully disentangle parameter space. Importantly, our findings also point to the relevance of warm inflationary dynamics: dissipative effects and the presence of a thermal bath can naturally enhance $n_s$, offering a pathway to reconcile theory with data without resorting to extreme parameter choices. Taken together, these results suggest that the $\alpha$-attractor models \cite{Kallosh:2013tua,Kallosh:2013hoa,Kallosh:2013maa,Kallosh:2013yoa,Kallosh:2014rga,Kallosh:2015lwa,Roest:2015qya,Linde:2016uec,Terada:2016nqg,Ueno:2016dim,Odintsov:2016vzz,Akrami:2017cir,Dimopoulos:2017zvq}, the $R^2$ inflationary framework \cite{Starobinsky:1980te}, and Higgs inflation with a strong nonminimal coupling to gravity \cite{Kaiser:1994vs,Bezrukov:2007ep,Bezrukov:2008ej}, possibly within the warm inflation paradigm, represents a theoretically robust and observationally viable alternative to conventional single-field models.

Warm inflation may suggest a preference for higher values of $n_s$, introducing tension with the simplest realizations of the models include Starobinsky $R^2$ model, as well as Higgs and Higgs-like scenarios. As pointed out in the literature, theses models are in tension with the latest observational data \cite{ACT:2025fju, ACT:2025tim}, it would be worthwhile to investigate the inclusion of dissipation of the inflaton field into the radiation bath. As noted in Ref.~\cite{Berera:2025vsu}, the warm inflation framework, with physically well-motivated dissipative terms, offers a consistent paradigm that successfully accounts for current observational data. There are some existing publications on the subject relevant to warm inflationary dynamics, including non-minimally coupled Peccei–Quinn inflation \cite{Yuennan:2024nje}, nonminimally-coupled Higgs inflation \cite{Eadkhong:2023ozb}, quantum-corrected self-interacting inflaton potential \cite{Samart:2021hgt}, and Higgs–Starobinsky inflation \cite{Samart:2021eph}. 

Finally, we explored particle production through parametric resonance within an inflationary framework based on quantum-corrected $\phi^4$ inflation with a non-minimal coupling to gravity. In this setup, the inflaton field $\phi$ interacts with an additional scalar $\chi$ via the coupling term $g^{2}\phi^{2}\chi^{2}$. In flat (Minkowski) spacetime, the resonance dynamics reduce to the well-known Mathieu equation. Our analysis shows that broad resonance regimes can naturally occur, making the mechanism highly efficient in this model. 


\end{document}